\documentclass[%
 reprint,
showpacs,
nofootinbib,
nobibnotes,
amssymb,
 aps,
]{revtex4-1}

\usepackage{graphicx}% Include figure files
\usepackage{dcolumn}% Align table columns on decimal point
\usepackage{bm}% bold math
\usepackage{subfig}
\usepackage{fullpage}

\usepackage{amsmath}
\usepackage{hyperref}
\usepackage{cleveref}
\usepackage{amsfonts}
\usepackage{amssymb}
\usepackage{algorithm}
\usepackage{algpseudocode}
\usepackage{float}
\usepackage{relsize}
\usepackage{epstopdf}

\begin{document}

\preprint{APS/123-QED}
%\title{A Fast Multipole Method and a Metropolis Method for Coarse-grained Brownian Dynamics Simulations of a DNA with Hydrodynamic Interactions}
\title{Efficient Brownian Dynamics Simulation of Single DNA with Hydrodynamic Interactions in Linear Flows}
%\title{A Fast Multipole Method for Coarse-grained Brownian Dynamics Simulations of a DNA with Hydrodynamic Interactions}% Force line breaks with \\
\thanks{Simulation of single DNA in linear flows}%

\author{Szu-Pei Fu}
 \email{sf47@njit.edu}
% \altaffiliation[Also at ]{Department of Mathematical Sciences and Center for Applied Mathematics and Statistics, \\
%New Jersey Institute of Technology, Newark, New Jersey 07102 USA}%Lines break automatically or can be forced with \\
\author{Yuan-Nan Young}%
 \email{yyoung@njit.edu}
\author{Shidong Jiang}
 \email{shidong.jiang@njit.edu}
\affiliation{%
Department of Mathematical Sciences and Center for Applied Mathematics and Statistics, \\
New Jersey Institute of Technology, Newark, New Jersey 07102 USA
}%


%\author{  }
%\affiliation{
%\affiliation{
 %Third institution, the second for Charlie Author
%\affiliation{%


\date{\today}% It is always \today, today,

\begin{abstract}
The coarse-grained molecular dynamics (MD) or Brownian dynamics (BD) simulation
is a particle-based approach that has been applied to a wide range of biological problems
that involve interactions with surrounding fluid molecules or the so-called hydrodynamic interactions (HIs).
In this paper, an efficient algorithm is proposed to simulate the motion of a single DNA molecule
in linear flows. The algorithm utilizes the integraing factor to cope with the effect of 
the linear flow of the surrounding fluid and applies the Metropolis method (MM) in 
 [N. Bou-Rabee, A. Donev, and E. Vanden-Eijnden, Multiscale Model. Simul. {\bf 12}, 781 (2014)]
to achieve more efficient BD simulation. Thus our method permits much larger time step size than previous 
methods while still maintaining the stability of the BD simulation, which is
advantageous for long-time BD simulation. Our numerical results on $\lambda$-DNA agree very well with
both experimental data and previous simulation results. Finally, when combined with fast algorithms
such as the fast multipole method which has nearly optimal complexity in the total number of beads, the 
resulting method is parallelizable, scalable to large systems, and stable for large time step size,
thus making the long-time large-scale BD simulation within practical reach.  This will be useful
for the study of membranes, long-chain molecules, and a large collection of molecules in the fluids.
%
%
%%%%%%%%%%%%%%%%%%%%%%%%%%%%%%%%%%%%%%%%%%%%%%%%%%%%%%%%%%%%%%
%
%
% biologically relevant problems the non-local hydrodynamic interactions (HIs) of macromolecules in 
%water are essential to the underlying physics for their dynamics. A lot of water %molecules need to 
%be included to capture the non-local HIs. As a result the simulations become numerically expensive 
%even for the coarse-grained dynamics of a small system over a short %time. The particular focuses 
%are on validating the improved coarse-grained BD schemes for a DNA macromolecule and the 
%kernel-free fast multipole method for computing the non-local %hydrodynamic interactions.
%
%
%
%\begin{description}
%\pacs{
%\end{description}
\end{abstract}

\maketitle

%\tableofcontents

\section{Introduction}
\label{sec:level1}
The dynamics of a single DNA or polymer macromolecule in fluid flow has been 
extensively investigated experimentally (\cite{Chu97,Chu99} and references therein), 
theoretically \cite{Ermak78,Marko95,DoiEdwards} and numerically \cite{Graham02,Schroeder04}. 
Bulk rheological experiments such as flow bi-refringence and light scattering measurements 
give inference 
of polymer conformation, orientation, and chain stretch in fluid flows. The advent of single 
molecule visualizations 
using fluorescence microscopy allows for the direct observation of complex dynamics of individual 
macromolecules 
in dilute solutions under shear, extensional, and general two-dimensional mixed flows 
\cite{Chu00,Larson03,Schroeder03,Chu99,Chu98}. These measurements
provide data for direct comparison against fully parametrized models of macromolecules, 
such as the 
bead-spring model for DNA with finite extensibility, excluded volume (EV) \cite{Prakash02} effects 
and hydrodynamic interactions (HI) \cite{Schroeder04}.
Brownian dynamics (BD) simulations of bead-spring and bead-rod models 
with free-draining assumption (no hydrodynamic interactions) 
give quantitative agreement with dynamical polymer behavior from single-molecule 
experiments \cite{Graham00,Larson99,Shaqfeh02}.

Following Ermak and McCammon \cite{Ermak78}, Schroeder {\it et al.} modeled the DNA 
macromolecule as a system of $N$ particles subject to
interparticle forces, fluctuating HI and EV forces \cite{Schroeder04}. 
They designed a semi-implicit predictor-corrector scheme for simulating the Brownian system,  
and illustrated how effects of HI and EV between monomers in a flexible polymer chain influence
both the equilibrium and non-equilibrium physical properties of DNA macromolecules 
\cite{Schroeder04}, consistent with the experimental observations.
The non-local HI between the DNA macromolecule and the surrounding fluid involves an integral
of hydrodynamic forces between a point and the rest of the macromolecule. 
Within the coarse-grained framework,
this integral is equivalent to a sum of all hydrodynamic forces between a bead and the rest of the system.
Here we adopt the Rotne-Prager-Yamakawa (RPY) tensor \cite{RP69} (i.e., the mobility tensor) for HI effects:
\begin{equation} 
\begin{aligned}[c]
\label{rpy1}
		D_{ij} = \frac{k_BT}{\zeta_{res}}I_{ij}  \textrm{,  if $i=j$}
\end{aligned}
\end{equation}
\begin{equation}
\begin{aligned}[c]
\label{rpy2}
		D_{ij} = \frac{k_BT}{8\pi\eta r_{ij}}\bigg[ \left(1+\frac{2a^2}{3r^2_{ij}}\right)I_{ij} + \left(1-\frac{2a^2}{r^2_{ij}}\right)\frac{\textbf{r}_{ij}\textbf{r}_{ij}}{r^2_{ij}} \bigg] \\
 \textrm{, if $i\neq j$, $r_{ij}\geq 2a$}
\end{aligned}	
\end{equation}
\begin{equation}
\begin{aligned}[c]
\label{rpy3}	
		D_{ij} = \frac{k_BT}{\zeta_{res}}\bigg[ \left(1-\frac{9r_{ij}}{32a}\right)I_{ij} + \frac{3\textbf{r}_{ij}\textbf{r}_{ij}}{32ar_{ij}} \bigg] \\
 \textrm{,  if $i\neq j$, $r_{ij}< 2a$}
\end{aligned}	
\end{equation}
where $D_{ij}$ is the mobility of bead $i$ due to bead $j$ in three dimensions, $I_{ij}$ the $3\times3$ identity matrix, and
$\zeta_{res}=6\pi\eta a$ is the bead resistivity with $\eta$ the solvent viscosity and
$a$ the radius of beads. 

%
%The (mobility) RYP tensor $D_{ij}$~\cite{RP69} was used for this sum, and 
%the calculation of HI is 
%

There are two main challenges for the long-time large-scale BD simultions with HI and EV effects.
First, the correlated random noises in the change of displacement vectors at each time step are proportional
to $\sqrt{\Delta t}$ with $\Delta t$ the time step size. This makes the design of high-order marching 
scheme very difficult and forces very small $\Delta t$ for many explicit or semi-implicit numerical
schemes in order to avoid the numerical instability. The problem becomes much more severe for 
long-time BD simulations
since it then requires a very large total number of time steps for the system to reach the desired state, 
which very often leads to weeks of simulation time even for one run. 

Second, the direct evaluation of 
the particle interaction at each time step requires $O(N^2)$ operations where $N$ is the total number of particles
in the system; and the generation of the correlated randome displacements requires $O(N^3)$ operations
if the standard Choleski factorization is used or $O(KN^2)$ if the Chebyshev spectral approximation is used
for computing the product of the matrix square root and an arbitrary vector (here $K$ is the condition number 
of the covariance matrix) (see, for example, \cite{Jiang13_MC}). 

To summarize, 
in order to efficiently utilize the BD simulation as a practical tool to study the properties of large systems, 
say, many polymers or a large collection of DNA molecules in a fluid, 
it is essential to address the following two questions: how to numerically
integrate the system with greater accuracy and better stability property which enables much large time step size? 
How to expedite the calculations of long-range particle interactions and associated correlated random effects 
in BD simulations with HI, especially for large $N$? 

For BD simulations near equilibrium, a Metropolis scheme for the temporal integration has been recently proposed 
\cite{BouRabee14,BouRabee14v2} for a Markov process whose generator is self-adjoint (with respect to a 
density function) to expedite simulations to reach equilibrium in a timely fashion. Under this scheme, 
stable and accurate BD simulations of DNA in a solvent are obtained using time step sizes that are orders 
of magnitude larger than those for predictor-corrector schemes \cite{Larson99,Graham02,Schroeder04}. However,
such a Metropolis scheme relies heavily on the self-adjointness of the Markov process generator 
for a quiescent flow.

In this work, we present an efficient algorithm for the 
simulations of the dynamics of DNA macromolecules under linear flows. Our method is based upon the 
Metropolis scheme developed in \cite{BouRabee14v2} for self-adjoint diffusions, which is applicable for the 
study of the DNA molecule to its equilibrium configurations in a quiescent flow. When a linear flow such as an 
extensional or a shear flow is present in the surrounding fluid, the diffusion process is not self-adjoint anymore. 
We first apply the method of integrating factors to recast the associated system of stochastic differential equations
(SDE) into a form such that the effect of the linear flow is taken into account by the integrating factor. 
We then modify the Metropolis scheme in \cite{BouRabee14v2} to update the displacements of beads which are the 
coarse-grained representation of the long chain DNA molecule. Our numerical experiments show that our scheme allows 
much greater time step size in the BD simulation and avoids the numerical instability. The numerical results
on the study of $\lambda$-DNA 
agree very well with the expermental data \cite{Chu98,Chu99} and previous simulation results 
\cite{Schroeder04}. Moreover, the total simulation time is significantly reduced in our methods as compared with 
the semi-implicit predictor-corrector scheme \cite{Schroeder04}. 

When the system involves a large
number of particles, as in the case of the study of lipid bilayer membranes, long chain polymers, or a large 
collection of DNA molecules, we observe that recent work in \cite{Jiang13_MC,Jiang13_JCP} reduces the 
computational cost of particle interactions from $O(N^2)$ to $O(N)$ and the cost of generating the correlated 
random displacements from $O(N^3)$ or $O(KN^2)$ to $O(KN)$, which leads to an essentially linear algorithm 
with respect to the total number of particles in the BD simulation. The method developed in 
\cite{Jiang13_MC,Jiang13_JCP} extends the original fast multipole method (FMM) \cite{fmm} to the case of the RPY 
tensor and combines it with the spectral Lanczos decomposition method (SLDM) to generate corelated random
vectors whose correlation is determined by the RPY tensor. We expect that long-time large-scale BD simulations
(with or without linear flows) for large systems are within practical reach when our modified Metropolis
scheme is combined with the method in \cite{Jiang13_MC,Jiang13_JCP}. Since the experimental or simulation 
results for large systems do not seem to appear in the literature, we will only present some numerical results
which indicate the linear scaling of methos in \cite{Jiang13_MC,Jiang13_JCP} and report the results on the 
BD simulation of large system on a later date.
 
This paper is organized as follows. 
In \cref{section:BDS} the formulation for the BD simulation is presented along with a discussion 
on the relevant physical parameters and forces. Section~\ref{section:NM} 
provides a detailed description of the 
numerical method used in this paper. In \cref{section:NR} we demonstrate the performance of our numerical 
scheme by comparing our numerical results with the experimental data \cite{Chu98,Chu99} and previous
simulation results \cite{Schroeder04} where the motion of a single DNA molecule in a quiescent, extensional,
or shear flow is studied and the DNA molecule is modeled via $29$ beads.
In \cref{section:FMM} we briefly discuss the extension of our method to the study of large systems by combining
it with the FMM for the RPY tensor and other fast algorithms.
Finally \cref{section:CD} contains a short conclusion and discussion for future work.
% 
%
%
%
%%%%%%%%%%%%%%%%%%%%%%%%%%%%%%%%%%%%%%%%%%%%%%%%%%%%%%%%%%%%
%
%
\section{Brownian Dynamic Simulation of a DNA molecule with HI}
\label{section:BDS}

The DNA or polymer macromolecule is coarse-grained into a system of $N$ beads described by the
Langevin equation \cite{Ermak78} with hydrodynamic interactions. 
The governing equation for the position vector ${\bf r}_i$ of the $i$th bead is 
\begin{equation}
\label{eq:main_eq}
	m_i\frac{d^2 {\bf r}_i}{dt^2} =  \sum_j\zeta_{ij}\cdot\left({\bf v}_j - \frac{d{\bf r}_j}{dt}\right) + {\bf F}_i + \sqrt{2}\sum_j\sigma_{ij}\cdot W_j,
\end{equation}
where $m_i$ is the mass of bead $i$, ${\bf v}_j$ is the solvent velocity, and $\zeta_{ij}$ is the friction 
coefficient tensor. The coefficient matrix $\sigma$ connects the thermal fluctuations of the particles through 
hydrodynamic interactions. In the Ermak-McCammon
model \cite{Ermak78}, it is related to $\zeta$ with
$\mathlarger{\zeta = \sigma^\top\sigma/k_BT}$, where $k_BT$ is the thermal energy. $W_j$ is the thermal fluctuation 
modeled as a Wiener process with mean $0$ and variance $dt$.
Thus, the RHS of \cref{eq:main_eq} is the total force acting on the bead $i$ including the drag force, 
total inter-particle force and the thermal fluctuating HI.

Ignoring the bead inertia, \cref{eq:main_eq} can be written as a first-order stochastic differential 
equation (SDE):
	\begin{equation}
	\begin{aligned}[c]
	\label{eq:inertialess_Leq}
		d{\bf r}_i  = \left(\kappa\cdot {\bf r}_i + \sum_{j=1}^N\frac{\partial D_{ij}}{\partial {\bf r}_j} + \sum_{j=1}^N\frac{D_{ij}\cdot {\bf F}_j}{k_BT}\right)dt \\
+ \sqrt{2}\sum^i_{j=1}\alpha_{ij}\cdot dW_j,
	\end{aligned}
	\end{equation}
where $\kappa$ is the transpose of the constant velocity gradient tensor of the
linear far-field flow velocity and ${\bf v}_i = \kappa\cdot {\bf r}_i$ (${\bf v}_j=0$ in a quiescent flow). 
The random Wiener process in the SDE $dW_j$ is related to $dt$ as: $dW_j=\sqrt{dt} {\bf n}_j$ where ${\bf n}_j$
is a random vector with the standard Gaussian distribution.

$D$ is the mobility tensor of size $3N\times 3N$ and for the N-bead chain the tensor $D$ is related to the 
thermal energy through the friction coefficient tensor $\zeta_{ij}$ as
$\sum_l \zeta_{il}D_{lj} = k_BT\delta_{ij}$. As in \cite{Ermak78,Schroeder04}, we use the RPY tensor for 
$D$. 

In the absence of external driving forces, the covariance between the bead displacements satisfy 
the following relation
\begin{equation}
	\langle d\textbf{r}_id\textbf{r}_j\rangle=2D_{ij}dt.
\end{equation} 
Hence, the coefficient matrix $\alpha$ is connected with $D$ via the formula 
$\mathlarger{D = \alpha^\top\alpha}$. We remark here that the choice of $\alpha$ is not unique and fast algorithms
for generating these correlated random displacements actually take advantage of this fact. Finally, we observe
that for the RPY tensor, $\sum_{j=1}\frac{\partial D_{ij}}{\partial {\bf r}_j}$ is always zero and 
\cref{eq:inertialess_Leq} is reduced to 
	\begin{equation}
%	\begin{aligned}[c]
	\label{sde1}
		d{\bf r}_i  = \left(\kappa\cdot {\bf r}_i + \sum_{j=1}^N\frac{D_{ij}\cdot {\bf F}_j}{k_BT}\right)dt 
+ \sqrt{2}\sum^i_{j=1}\alpha_{ij}\cdot dW_j.
%	\end{aligned}
	\end{equation}

\subsection{Nondimensionalization of the SDE \eqref{sde1}}

The bead-spring chain model is widely used for BD simulations of a DNA molecule. In the bead-spring chain model,
the DNA molecule is represented as a chain of $N$ beads of radius $a$ with adjacent beads connected by a spring.
Each spring contains $N_{k,s}$ Kuhn steps of length $b_k$. So the maximum length of each spring is $N_{k,s}b_k$,
and the characteristic contour length of the double stranded DNA molecule $L$ is approximately $(N-1)N_{k,s}b_k$
as the size of each bead is much smaller than the length of each spring and thus neglected.
We denote the Hookean spring constant
by $H$. The characteristic length $l_s$ is chosen to be $l_s=\sqrt{k_BT/H}$ and the characteristic time
$t_s$ is chosen to be $t_s=\zeta_{res}/4H$, where $\zeta_{res}$ is the bead resistivity appeared in the RPY tensor 
\eqref{rpy3}. We scale the length and time by $l_s$ and $t_s$, respectively and nondimensionalize \cref{sde1}
into the following dimensionless form:
	\begin{equation}
	\label{eq:SDE}
		d{\bf r}_i = \left(\kappa\cdot {\bf r}_i + \sum^N_{j=1}D_{ij}\cdot {\bf F}_j \right)dt + \sqrt{2}\sum^i_{j=1}\alpha_{ij}\cdot dW_j,
	\end{equation}
Here with a slight abuse of notation, we have used the same notation to denote all corresponding 
dimensionless quantities.

\subsection{Choices of the Velocity Gradient Tensor $\kappa$}
We now specify the velocity gradient tensor $\kappa$ in \cref{eq:SDE}
and restrict our attention to the following two linear planar flows. The first one is the extensional
flow where $v_x=\dot{\epsilon}x, v_y=-\dot{\epsilon}y$ with $\dot{\epsilon}$ the extension rate.
The second is the shear flow where $v_x=\dot{\gamma}y, v_y=0$ with $\dot{\gamma}$ the shear rate. 
We define the Peclet number  
$Pe=\dot{\epsilon}\zeta/4H$ for the extensional flow and $Pe=\dot{\gamma}\zeta/4H$ for the shear flow, respectively.
Then the dimensionless velocity gradient tensor $\kappa$ in \cref{eq:SDE} is given by the following formulas:
\begin{equation}
\begin{array}{cc}
\kappa_{ext} =
 \begin{pmatrix}
  Pe & 0 & 0 \\
  0 & -Pe & 0 \\
  0  & 0  & 0  \\
 \end{pmatrix}, \;\;\; &

\kappa_{shear} =
 \begin{pmatrix}
  0 & Pe & 0 \\
   0 & 0 & 0 \\
  0  & 0  & 0  \\
 \end{pmatrix}.
\end{array}
\end{equation}
Here $\kappa=\kappa_{ext}$ for the extensional flow and $\kappa=\kappa_{shear}$ for the shear flow.

%%%%%%%%%%%%%%%%%%%%%%%%%%%%%%%%%%%%%%%%%%%%%%%%%%%%%%%%%%%%%%%%%%%%%%%%%%%%%%%%

\subsection{Specification of the Forcing Term ${\bf F_i}$}
The force ${\bf F_j}$ in \cref{eq:SDE} contains two parts: the force exerted by the connected springs and 
the force due to the finite size of the beads. We adopt the  
Marko-Siggia's wormlike chain (WLC) spring law \cite{Marko95}
to model the spring force between beads. 
In the WLC model, the dimensionless spring force acting on the $i$th bead by the $i$th spring is
	\begin{equation}
	\label{eq:sp_force_dl}
		{\bf F}^s_i = \sqrt{\frac{N_{k,s}}{3}}
\left[\frac12\frac{1}{\left(1-\frac{Q_i}{Q_0}\right)^2}-\frac12+\frac{2Q_i}{Q_0}\right]\frac{{\bf Q}_i}{Q_i},
	\end{equation}
where $i=1,\dots,N-1$, ${\bf Q}_i={\bf r}_{i+1}-{\bf r}_i$ is the distance vector between bead ${\bf r}_{i+1}$ 
and ${\bf r}_i$, $Q_i$ is the length of ${\bf Q}_i$, and $Q_0$ is the maximum distance between these two beads.
Since all interior beads are connected with two springs from two sides, the net entropic spring force
acting on the $i$th bead is
\begin{equation}
	\textbf{F}^{\text{entropy}}_i 
=  {\bf F}^s_i - {\bf F}^s_{i-1},  \;\;\;{\bf F}^s_0={\bf F}^s_N=0,
\end{equation}
with $i=1,\dots,N$.
For later use, we also record the potential for the $i$th spring below 
	\begin{equation}
			U_{WLC}({\bf Q}_i) = \frac{1}{2}\sqrt{\frac{N_{k,s}}{3}}\bigg(\frac{Q_0^2}{Q_0-Q} - Q + \frac{2Q^2}{Q_0}\bigg).
	\end{equation}

For the force due to the finite size of the beads, we adopt the excluded volume force in \cite{Prakash02,Schroeder04}
given by the formula
	\begin{equation}
		{\bf F}^{EV}_{i} 
= -\sum^N_{j=1,i\ne j}\ \frac{9\sqrt{3}z}{2}
\text{exp}\left(-\frac{3r^2_{ij}}{2}\right){\bf r}_{ij}
	\end{equation}
where $z = \left(\frac{1}{2\pi}\right)^{3/2}\tilde{v}N^2_{k,s}$, and $\tilde{v}=2ab_k^2/l_s^3$ 
is the dimensionless excluded volume parameter. And the excluded volume potential between bead $i$ and bead $j$ is
given by  
	\begin{equation}
		U^{EV}_{ij} =\frac{3\sqrt{3}z}{2} \text{exp}\left(-\frac{3r^2_{ij}}{2}\right).
	\end{equation}
Finally, the total force acting on bead $i$ is the sum of the spring forces and the excluded volume forces, that
is,
	\begin{equation}
		{\bf F}_i= {\bf F}^{\text{entropy}}_i + {\bf F}^{EV}_{i} .
	\end{equation}

%%%%%%%%%%%%%%%%%%%%%%%%%%%%%%%%%%%%%%%%%%%%%%%%%%%%%%%%%%%%%%%%%%%%%%%%%%%%%%%%

\section{Numerical Algorithm for BD Simulations in Linear Flows}
%\section{Numerical Methods}
\label{section:NM}
%\subsection{Modified Self-adjoint Metropolis Integrator for BD simulations}
%\label{subsec:Metropolis} 

In the past, a semi-implicit predictor-corrector scheme \cite{Larson03,Schroeder04,Shaqfeh02} 
is often used for the temporal integration in BD simulations. A major problem associated with that scheme
is that a very small time step size has to be used in order to avoid the numerical instability,
which leads to an excessively large number of time steps and a very long total simulation time.
Recently, a Metropolis integrator has been developed to 
integrate the self-adjoint diffusion equations \cite{BouRabee14v2} for BD simulations in a quiescent flow. 

Here we extend the algorithm in \cite{BouRabee14v2} to study BD simulations in linear flows. We first 
introduce an integrating factor $e^{-\kappa t}$ and rewrite \cref{eq:SDE} as follows:

\begin{equation}
\label{eq:stochastic_Leq_ext}
		d\big(e^{-\kappa t}{\bf r}_i\big) = \text{e}^{-\kappa t}\bigg[\sum\limits^N_{j=1}D_{ij}\cdot {\bf F}_jdt + \sqrt{2}\sum^i_{j=1}\alpha_{ij}\cdot dW_j\bigg].
\end{equation}

Similar to the algorithm in \cite{BouRabee14v2}, we now update the position vector as follows:

1. Compute the vector $\hat{\bf r}_i^{n+1}$ as follows:
\begin{equation}
\begin{aligned}
	\tilde{\bf r}_i^{n+1} &= {\bf r}_i^{n} + \sqrt{\frac{dt}{2}}B({\bf r}_i^{n})dW_i,\\
	\hat{\bf r}_i^{n+1} &= \tilde{\bf r}_i^{n+1} + G(\tilde{\bf r}_i^{n+1})\Delta t + (\tilde{\bf r}_i^{n+1} - {\bf r}_i^{n}),
\end{aligned}
\end{equation}

where the functions $G$ and $B$ are defined by the following formulas:

\begin{equation}
\begin{aligned}
	{\bf x}_1  & = {\bf x} + \frac23 D({\bf x}){\bf F}({\bf x})\Delta t,\\
	G({\bf x}) & = \frac58 D({\bf x}){\bf F}({\bf x}) - \frac38 D({\bf x}){\bf F}({\bf x}_1) \\
			  & - \frac38 D({\bf x}_1){\bf F}({\bf x}) + \frac98 D({\bf x}_1){\bf F}({\bf x}_1), \\
\end{aligned}
\end{equation}

\begin{equation}
\begin{aligned}
	{\bf x}_2  & = {\bf x} - \frac23 D({\bf x}){\bf F}({\bf x})\Delta t,\\
	B({\bf x}) B({\bf x})^\top & = \frac14 D({\bf x}) +  \frac34 D({\bf x}_2).
\end{aligned}
\end{equation}

2. Calculate the acceptance probability $\alpha$ as follows:

\begin{equation}
%\small
\begin{aligned}
	& \alpha({\bf r}_i^{n},\hat{\bf r}_i^{n+1}) = \min\bigg(1,\bigg.\\
& \bigg.
C\exp\bigg[ -\frac{|d\hat{W}|^2}{2} + \frac{|dW|^2}{2} - U(\hat{\bf r}_i^{n+1}) + U({\bf r}_i^{n})\bigg] \bigg),
\end{aligned}
\end{equation}
where $C=\det B({\bf r}_i^n)
/\det B(\hat{\bf r}_i^{n+1})$, $U=U_{WLC}+U^{EV}$ is the total potential energy, and
$d\hat{W}_i$ is obtained via the formula
\begin{equation}
	B(\hat{\bf r}_i^{n+1})
d\hat{W}_i = B({\bf r}_i^n)dW_i + \sqrt{2\Delta t}G(\tilde{\bf r}_i^{n+1}).
\end{equation}

3. Generate a Bernoulli random number $\gamma$, that is, generate a 
uniformly distributed random number $\beta$ on $[0,1]$ and set $\gamma$ to $1$ if $\beta\leq\alpha$ and 
$0$ otherwise.

4. Compute the updated position vector at time $t=t_{n+1}$ by the formula

\begin{equation}
        \label{rnp1}
	{\bf r}_i^{n+1} = \gamma A \hat{\bf r}_i^{n+1} + (1-\gamma) {\bf r}_i^{n}
\end{equation}

with the matrix $A=A_{ext}$ or $A=A_{shear}$ for the extensional or shear flow, respectively. Here
$A_{ext}$ and $A_{shear}$ are given by the formulas: 
\begin{equation}
A_{ext} =
 \begin{pmatrix}
  \text{e}^{-Pe\Delta t} & 0 & 0 \\
  0 &   \text{e}^{Pe\Delta t} & 0 \\
  0  & 0  & 1  \\
 \end{pmatrix}, 
\end{equation}
\begin{equation}
A_{shear} =
 \begin{pmatrix}
  1 & -Pe\Delta t & 0 \\
   0 & 1 & 0 \\
  0  & 0  & 1  \\
 \end{pmatrix}.
\end{equation}

In other words, the position vector will be updated only if the Bernoulli random number $\gamma$ is equal to $1$.
This is the essence of the Metropolis algorithm for Monte-Carlo simulations.

%%%%%%%%%%%%%%%%%%%%%%%%%%%%%%%%%%%%%%%%%%%%%%%%%%%%%%%%%%%%%%%%%%%%%%%%%%%%%%%%

\section{Numerical Results}
\label{section:NR}

Common measures of the ``stretch" of a DNA molecule under flow are the molecular fractional 
extension ($\hat x$ is the unit vector in the x direction)
\begin{equation}
\label{eq:extension}
X \equiv \max\limits_i({\bf r}_i\cdot \hat x) - \min\limits_i({\bf r}_i\cdot\hat x),
\end{equation}
and its ensemble average $\langle X \rangle \equiv \frac{1}{M}\sum X$, where $M$ is the total 
number of experiments (or simulations).
Here we first compare the transient fractional extensions of a $\lambda$-DNA between the 
experimental data, 
semi-explicit numerical simulations \cite{Schroeder04}, and our Metropolis scheme simulations. 
The initial DNA configurations in these simulations are the equilibrium DNA configurations in the 
absence of flow from the Metropolis scheme. 

For the purpose of comparison, we use the same values of physical and model parameters as 
in \cite{Schroeder04}. That is, the viscosity $\eta$ 
of solvent is $8.4$ $cP$($=mPa\cdot s$) and the relaxation time $\tau$ is $21.0$ seconds. The 
$\lambda$-DNA is modeled with $N=29$ beads of radius $a=0.101\ \mu m$ connected by $28$ springs, where 
each spring has $N_{k,s}=40$ Kuhn steps of size $b_k=0.132$ $\mu m$ and  
the contour length $L$ is $150$ $\mu m$. Finally, 
the excluded volume parameter $v = 0.0034$ $\mu m^3$.

%This parameter set also implies that the hydrodynamic interaction parameter $h^*=0.12$. 

To mimic the experimental configurations, it is essential \cite{Schroeder04,SchroederPrivateComm} 
to first simulate the DNA molecule 
to its equilibrium in a quiescent flow, i.e.,
$\kappa\cdot{\bf r}_i=0$ in \cref{eq:SDE}, which is now a
self-adjoint stochastic differential equation that
can be efficiently solved to an equilibrium state using the Metropolis scheme in \cref{section:NM}.
At the beginning of the no-flow simulations, beads are equally spaced on the x-axis.
The Metropolis scheme allows for relatively large time step $\triangle t$
(an order of magnitude larger), 
consequently saving a significant amount of computation time for running 
no-flow simulations
compared to the semi-implicit predictor-corrector scheme in \cite{Schroeder04}.
The flow-free simulation is continued until an equilibrium configuration is reached, which is
often 10-20 relaxation times $(\tau)$.
After the equilibrium is reached for a DNA in a quiescent flow, we then turn flow on in the simulations 
and sum up $d{\bf r}_i$ to obtain the updated configuration and the mean fractional 
extension of a DNA molecule under linear flow. 
\begin{figure}
\begin{center}
\begin{tabular}{c}
\includegraphics[width=1.0\linewidth]{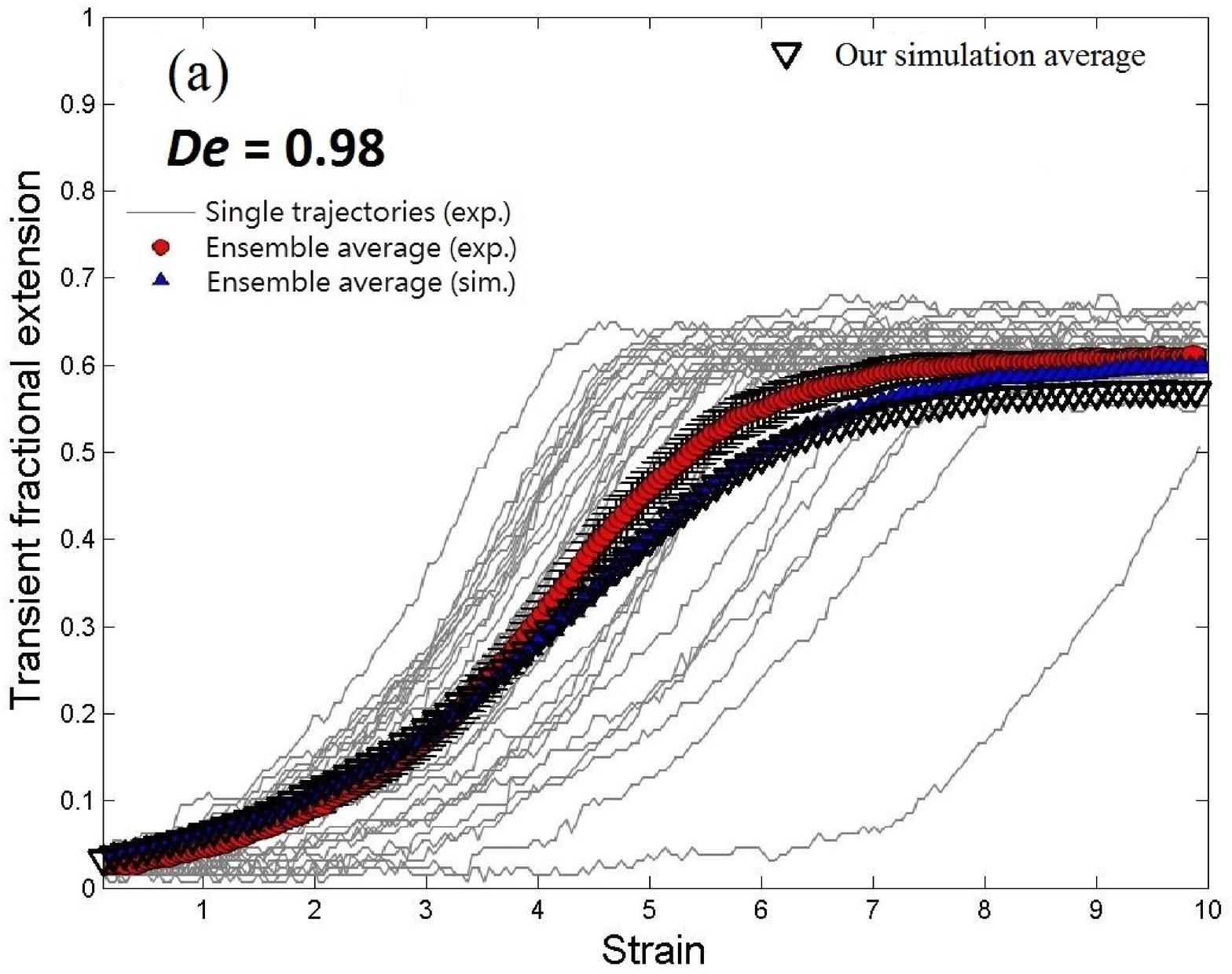} \\
\includegraphics[width=1.0\linewidth]{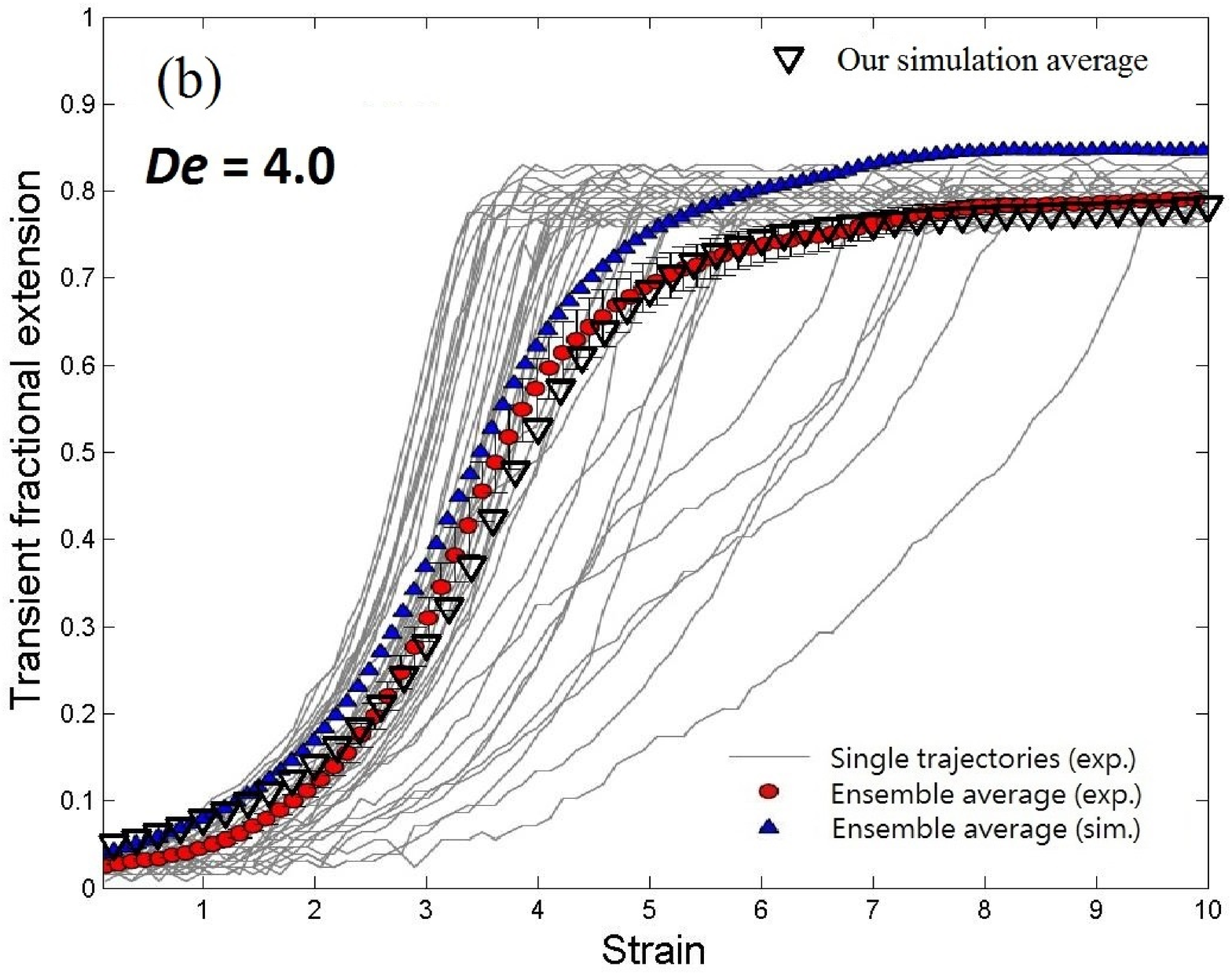} 
\end{tabular}
\caption{Transient fractional extension for a 7-lambda ($L=150$  $\mu m$) DNA in a planar extensional flow. 
$60$ trajectories from simulations are used for ensemble average.
%$De=0.98$  and $\triangle t \approx 10^{-4}$ sec/$t_s\sim 10^{-3}$ for panel (a), and
%$De=4$ and $\triangle t \approx 10^{-4}$ sec/$\epsilon\sim 5\times10^{-4}$ for panel (b).
}
\label{fig:2De_metro}
\end{center}
\end{figure}

The transient fractional extension from these simulations is summarized in
Figure~\ref{fig:2De_metro}, which shows two sets of comparison for Deborah 
number $De=0.98 \;(\dot\epsilon\approx0.0467 \text{ sec}^{-1})$ and $De=4.0 \;(\dot\epsilon\approx0.1905\text{ sec}^{-1})$ 
for panels (a) and (b), respectively. Figure~\ref{fig:2De_metro} 
is simulated by using the modified Metropolis integrator scheme with an integrating factor 
(\cref{eq:SDE} in \cref{section:NM}, $\kappa=\kappa_{ext}$). 
Thin curves are individual trajectories from experiments, 
filled circles are the ensemble average from experiments, filled triangles are ensemble average
from Schroeder {\it et al.} \cite{Schroeder04}, and our results are the empty triangles.
We observe that, in both panels, our results are in good agreement with
the experiment results.
However, our simulations are orders of magnitude more efficient
because a time-step $\Delta t=10^{-4}\tau =2.1\times10^{-3}\text{ sec}$ is used for 
results in panels (a), 
and $\Delta t=10^{-3}/\dot\epsilon =5.25\times10^{-3} \text{ sec}$ is used for panel (b).
In comparison, a much smaller time step 
%$\Delta t\approx \dot\epsilon\times 10^{-6}\approx 2\times 10^{-7} \text{ sec}$ 
for $De=4.0$  and $De=0.98$ cases are necessary 
for the predictor-corrector scheme \cite{Schroeder04,SchroederPrivateComm}.

\begin{figure} 
\begin{center}
\begin{tabular}{c}
\includegraphics[width=1.0\linewidth]{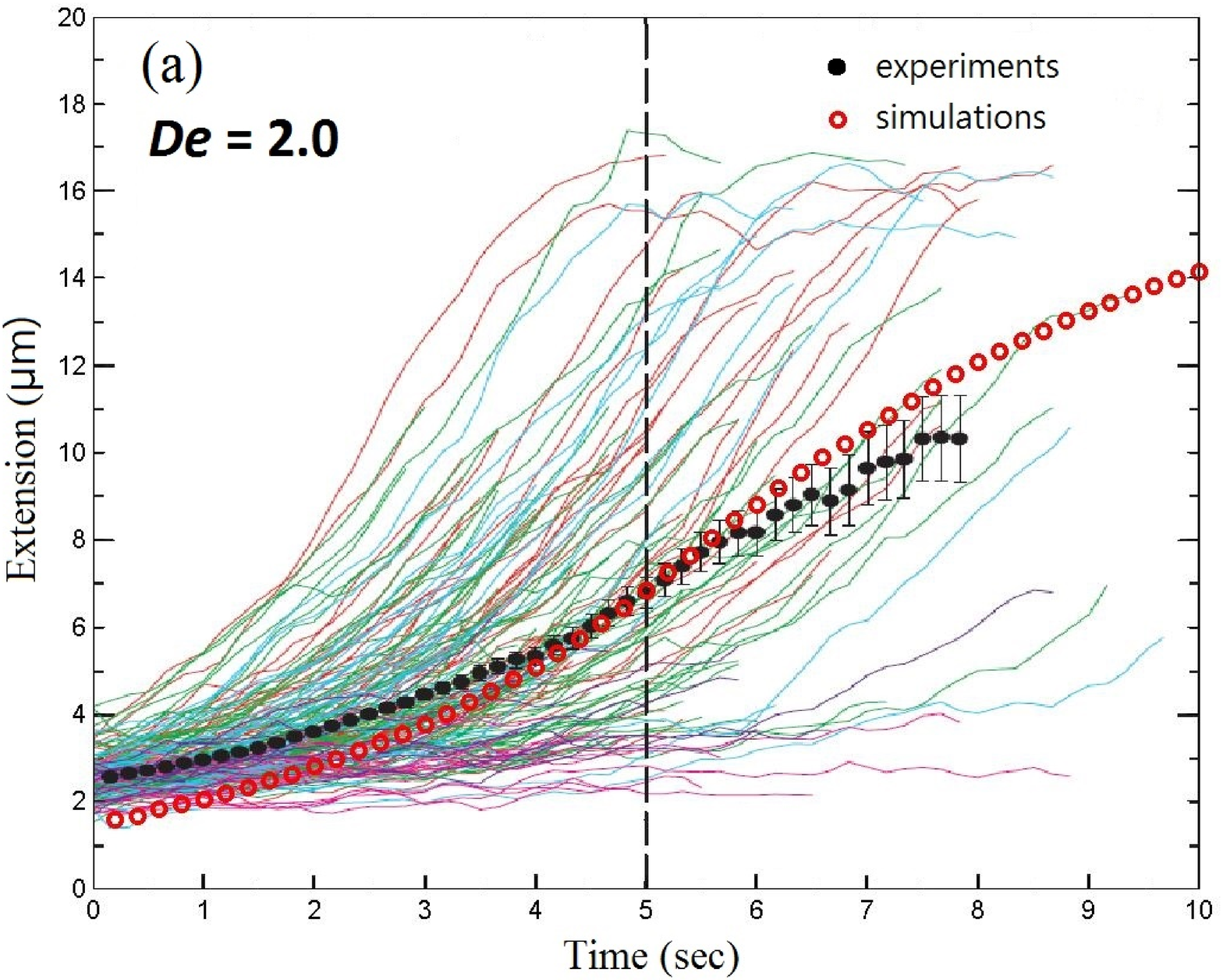} \\
\includegraphics[width=1.0\linewidth]{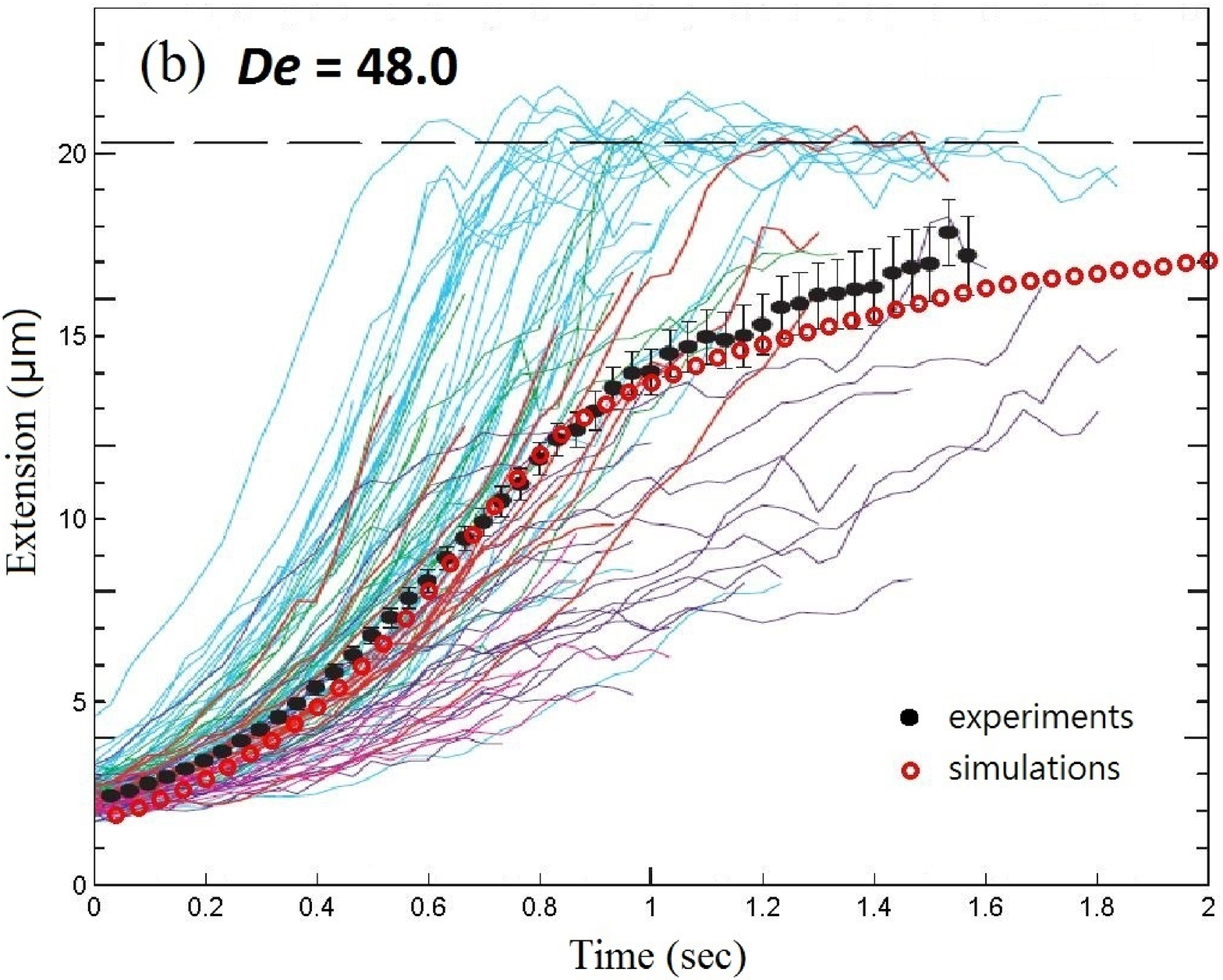}
\end{tabular}
\caption{Comparison between experiments \cite{Chu98} (thin curves for individual trajectories
and filled circles for the average) and 
our numerical simulations (empty circles).
 The vertical dashed line in figure $2(a)$ shows the point below which continuous data could not be
collected in some experiments. The horizontal dashed line in figure 2(b) shows the steady-state of the stretched
 $\sim22\ \mu m$ $\lambda$-DNA.
 %$\Delta t=10^{-3}\text{ sec}$ is used with $De=2.0$  for panel (a), and
 %$De=48.0$ for panel (b).
 }
\label{fig:Graham02}
\end{center}
\end{figure}

Similar comparison of a single DNA molecule in a planar extensional flow between
experiment and simulations are also conducted in \cite{Graham02}. 
%Based on the method
%proposed by Fixman \cite{Fixman86} for computing $\sqrt{D}$, 
%a first-order semi-implicit time integration scheme is
%developed \cite{Graham00} and the full algorithm scales roughly as
%$O(N^{2.25})$. Jendrejack {\it et al.} also reported substantial computational savings 
%over the standard Cholesky decomposition (which is used in \cite{Schroeder04}.)
%Here we perform the same simulations of a $21$ $\mu m$ DNA molecule in a planar extensional
%flow to compare against the simulation results in \cite{Graham02}. 
%Note that in our FMM method
%the algorithm scales as $O(N)$ for a three-dimensional long chain molecule.
Figure~\ref{fig:Graham02} compares our %FMM 
results against those from~\cite{Chu98}
for a $21$ $\mu m$ DNA molecule in an extensional flow with
$N=11$, $b_k=0.106 \text{ } \mu m$, $N_{k,s}=19.8$, $a=0.077 \text{ }\mu m$,
$23^{\circ}\mathrm{C}$ for the temperature and $v=0.0012\text{ } \mu m^3$.
Figure~\ref{fig:Graham02}(a) is for $De=2.0$, $\dot{\epsilon}=0.5\text{ sec}^{-1}$, 
$\tau=4.1 \text{ sec}$ and $\eta=43.3 \text{ cP}$. 
Figure~\ref{fig:Graham02}(b) is for $De=48.0$, $\dot{\epsilon}=2.8\text{ sec}^{-1}$, 
$\tau=17.3 \text{ sec}$ and $\eta=182 \text{ cP}$. 
The thin curves are trajectories from experiments~\cite{Chu98}, 
filled circles are the ensemble average of experimental results,
and empty circles are the ensemble average from our modified 
Metropolis integrator simulations.
For $De=2.0$ (panel (a)) our average is almost identical to the simulation average from \cite{Graham02} (bottom panel of their figure 2).
For $De=48.0$ (panel (b)),
it is clear that our simulation results are in better agreement with experimental results 
than those from Jendrejack {\it et al.} \cite{Graham02}. 
In these Metropolis integrator simulations $\Delta t=10^{-3} \text{ sec}$ for both 
$De=2.0$ in panel (a) and $De=48.0$ in panel (b).
Even though this time step is slightly smaller % larger ?? % 
than those used in \cite{Graham02}, 
our Metropolis algorithm with the integrating factor is second-order accurate \cite{BouRabee14,BouRabee14v2} and no matrix inversion is needed. In \cref{section:FMM}
we describe how our numerical algorithm can be further improved by
efficiently calculating the HI using FMM when the system size is large.

\begin{figure}
        \begin{center}
        \includegraphics[keepaspectratio=true,scale=0.2]{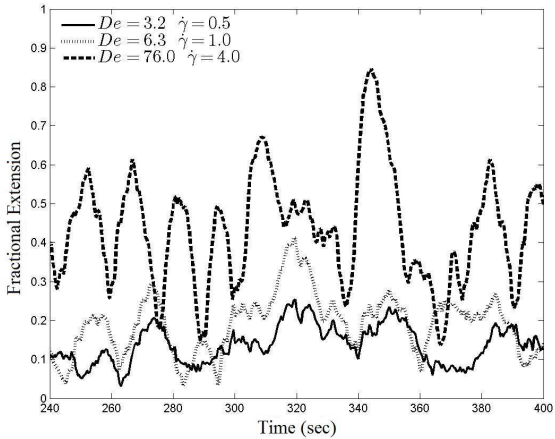}
       \end{center}
        \caption{Fractional extensions for different $De$ and shear rate $\dot\gamma$ 
        from our simulations with $(De,\dot\gamma) = (3.2, 0.5)$, $(6.3, 1.0)$, $(76.0, 4.0)$, 
        respectively. 
The relaxation time $\tau$ is $6.3\text{ sec}$ for the first two cases and 
$19.0 \text{ sec}$ for the third case. 
The time steps are: $\Delta t= 10^{-3}\ \text{sec}$ for $De=3.2$, 
$\Delta t= 5\times10^{-4}\text{ sec}$ for $De=6.3$, and 
$\Delta t= 2.5\times10^{-4}\text{ sec}$ for $De=76.0$. 
} 
        \label{fig:extension_vs_time}
\end{figure}
Next we compare the mean fractional extension of a DNA molecule 
against experiments \cite{Chu99} and Jendrejack {\it et al.}'s simulations \cite{Graham02}. 
The physical parameters in the experiments  are \cite{Chu99}: bead radius $a=0.077$ $\mu m$, and temperature is fixed at 
$20^{\circ}\mathrm{C}$. 
Two viscosities are considered in the experiments, 
$\eta=60 \textrm{ cP}$  and $220 \textrm{ cP}$ for the shear flow cases, while only
$\eta=60 \textrm{ cP}$ is used for the case of extensional flow (based on the experiments
in \cite{Chu99}).
For the corresponding simulations in \cite{Graham02} the
number of beads is $11$, Kuhn step size $b_k=0.106$ $ \mu m$, the number of springs 
per Kuhn step $N_{ks}=21$,
and the contour length $L=22$ $ \mu m$. 
\begin{figure}
		\centering
		\includegraphics[width=1\linewidth]{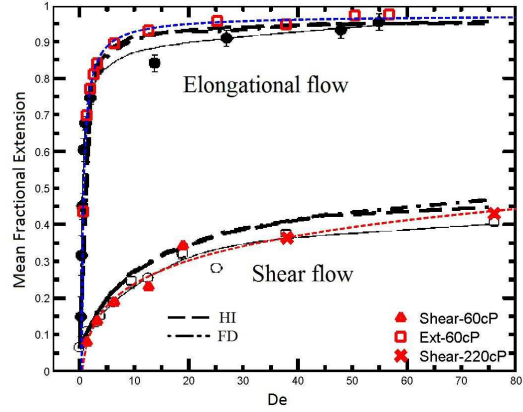}
		\caption{Mean fractional extensions for shear flow and extensional flow. Experimental data \cite{Chu99} are symbols with error bars, bead model with and without HI (see legend) are from 
		\cite{Graham02} and our results as red symbols (empty circles for the extensional flow; triangles and crosses for the shear flow). 
}
		\label{fig:extension_shear}
\end{figure}

Figure~\ref{fig:extension_vs_time} shows the fractional extension versus time for three cases: $De=3.2,\ 6.3,\ \text{and } 76.0$. 
As expected, larger mean extension of the DNA molecule is expected at a higher shear rate.
From these results the mean fractional extension is computed by taking the averages over
a long duration.

Figure~\ref{fig:extension_shear} shows the comparison of mean fractional extension 
between experiments \cite{Chu99}, Jendrejack {\it et al.}'s simulations \cite{Graham02} 
and our simulations. Experimental data are shown in filled dark disks for the extensional flow and 
dark circles for the shear flow, and the thin solid curves are their best fits.
Simulation results from \cite{Graham02} are thick dashed (with HI) and dash-dotted (without HI, or free-draining (FD)) curves.
Our simulation results are shown using the red symbols in the legends, 
and their best fits are the thin dashed curves.
It is clear that our results agree well with experimental data for the shear flow cases. 
For the extensional flow cases, our results agree better with simulation results from \cite{Graham02}  for all values of De.
At larger De (De $\ge 40$), all three agree well for the extensional flow cases.

%%%%%%%%%%%%%%%%%%%%%%%%%%%%%%%%%%%%%%%%%%%%%%%%%%%%%%%%%%%%%%%%%%%%%%%%%%%%%%%%

\section{Extension to Large Systems}
\label{section:FMM}
In the numerical algorithm described in \cref{section:NM}, the RPY tensor $D$ is constructed explicitly, 
the matrix vector product $D {\bf F}$ is computed directly, the uppertriangular matrix $B$ is obtained by the Cholesky
decomposition with its determinant simply the product of its diagonal entries. This is affordable
for the numerical experiments presented in \cref{section:NR} since the number of beads $N$ is only $29$.
However, for large systems, say, $N>1000$, the computational cost of these standard direct algorithms 
becomes prohibitively expensive since the matrix vector product $D{\bf F}$ requires $O(N^2)$ operations, 
the Cholesky factorization requires $O(N^3)$ operations, and each BD simulation often requires more than $10^5$ time
steps. Thus, fast algorithms become a necessity in order to make long-time large-scale BD simulations 
practical.

As mentioned in \cref{sec:level1}, recently a fast multipole method for the RPY tensor (RPYFMM) has been developed
in \cite{Jiang13_JCP}. The fundamental observation in \cite{Jiang13_JCP} is that the RPY tensor
can be decomposed as follows:
\begin{eqnarray}
\label{eq:rewrite_D}
%\begin{aligned}
%		D_{ij} &=  C_1\bigg(\frac{\delta_{ij}}{|\textbf{x}-\textbf{y}|}+\frac{(x_i-y_i)(x_j-y_j)}{|\textbf{x}-\textbf{y}|^3}\bigg)\\ 
%&+ C_2\bigg(\frac{\delta_{ij}}{|\textbf{x}-\textbf{y}|^3}-\frac{3(x_i-y_i)(x_j-y_j)}{|\textbf{x}-\textbf{y}|^5}\bigg) \\
			D_{ij}&=&  C_1\left[\frac{\delta_{ij}}{|\textbf{x}-\textbf{y}|}-(x_j-y_j)\frac{\partial}{\partial x_i}\frac{1}{|\textbf{x}-\textbf{y}|}\right] \\
&&+ C_2\frac{\partial}{\partial x_i}\frac{x_j-y_j}{|\textbf{x}-\textbf{y}|^3},\nonumber
%\end{aligned}
\end{eqnarray}
where $C_1=\frac{k_BT}{8\pi\eta}, C_2=\frac{k_BTa^2}{12\pi\eta}$.

With this decomposition, the matrix vector product $D {\bf v}$ for a given vector ${\bf v}$ can be interpreted
as a linear combination of four harmonic sums with suitably chosen source charges and dipoles. In other words,
the matrix vector product $D {\bf v}$ can be evaluated by four calls of the classical FMM for Coulomb interactions
in three dimensions \cite{cheng}. Thus, the RPYFMM avoids the explicit construction of the RPY tensor and reduces the computational cost of $D {\bf v}$ to $O(N)$ in both CPU time and memory storage. 

We observe further that the Cholesky factor $B$ of the RPY tensor $D$ can be replaced by any matrix $C$ which 
satisfies the same matrix equation $C C^{\top}=D$ (note that there are actually infinitely many matrices 
satisfying this matrix equation, see, for example, \cite{Jiang13_MC} for details). Indeed, \cite{Jiang13_JCP}
also proposed to replace the Cholesky factor $B$ by $\sqrt{D}$ and compute $\sqrt{D} {\bf v}$ by combining
the classical Spectral Lanczos Decomposition Method (SLDM) with the RPYFMM. The resulting algorithm has 
$O(\kappa N)$ complexity with $\kappa$ the condition number of the RPY tensor $D$. We remark here that 
for most BD simulations with HIs, the beads do not overlap with each other due to the EV force and our numerical
experiments show that the condition number of the RPY tensor in this case is fairly low. This indicates 
that the RPYFMM-SLDM method is essentially a linear algorithm for computing $\sqrt{D} {\bf v}$. 
The timing results presented in Table~\ref{table1} and Table~\ref{table2} clearly demonstrate of 
linear scaling of the RPYFMM and RPYFMM-SLDM methods.

\begin{table}[h!]
\begin{center}
    \begin{tabular}{  r  r  l  l }
    \hline
    N  & $T_{RPYFMM}$ & $T_{Direct}$ & $E_{RPYFMM}$\\ \hline
    		1,000 & 0.20897 & 0.31495 & 1.6008e-02\\ 
		10,000 & 1.6058 & 30.6643 & 5.5339e-02\\
		100,000 & 16.172 & 2738.48 & 8.3803e-02\\
		1,000,000 & 160.24 &  271009.4 & 1.1603e-01\\
    \hline

    \end{tabular}
\end{center}
\caption{Timing results (sec) for computing $T = D\textbf{v}$ by RPYFMM.}
\label{table1}
\end{table}

\begin{table}[h!]
\begin{center}
    \begin{tabular}{  r l  l  l }
    \hline
    N & m & $T_{SLDM}$ & $E_{relative}$\\ \hline
    		1,000 & 4 & 0.54192 &  6.21032e-06   \\ 
		10,000 & 4 & 9.03360 &  6.24604e-04  \\
		100,000 &  6 & 111.80 &  7.92857e-04   \\
		1,000,000  & 12 & 2180.8 & 2.91239e-04  \\
    \hline

    \end{tabular}
\end{center}
\caption{Timing results (sec) for computing $T = \sqrt{D}\textbf{v}$ by RPYFMM-SLDM.}
\label{table2}
\end{table}

Finally, we would like to remark here that recent developments in the fast multipole methods and fast
direct solvers also enable a linear algorithm for computing the determinant of a matrix with certain 
hierachical low rank structure \cite{kenho1,kenho2,siba}. By incorporating all these fast algorithms
into our current numerical scheme, we obtain a numerical algorithm which is stable even for relatively
large time step size and scales linearly with respect to the number of particals (or beads) in the system. 

%%%%%%%%%%%%%%%%%%%%%%%%%%%%%%%%%%%%%%%%%%%%%%%%%%%%%%%%%%%%

\section{Conclusion and Discussion}
\label{section:CD}

%We have presented that our method works for the cases of linear flow, our numerical simulations show that the time step can be chosen 1 order larger(or better) than the time step in %predictor-corrector scheme. The main contribution is to save the computational cost. For larger macromolecules, such as $1300\mu m$ $\lambda$-DNA, the time expense on constructing %the RPY tensor will be huge since the number of beads $N$ increases.  

We have extended the Metropolis integrator in \cite{BouRabee14v2} to study BD simulations 
with HIs in linear flows. The method utilizes the integraing factor to absorb the effect of 
the linear flow and permits much larger time step sizes for BD simulations with HIs in linear
flows. We have applied our method to study the 
fractional stretch and the mean stretch of a single $\lambda$-DNA molecule in planar linear 
flows. Our numerical results agree very well with  
experimental data \cite{Chu98,Chu99} and other simulation results \cite{Schroeder04}
in the literature.  

We have also discussed the extension of our method to large systems in \cref{section:FMM}. By
incorporating the RPYFMM and other fast algorithms into the scheme, the resulting algorithm admits
large time step sizes and has nearly optimal complexity (i.e., $O(N)$ or $O(N\log N)$)
in the number of particles in the system. 
Thus, even though many of these fast algorithms have a large prefactor 
(say, $C\geq 1000$) in front of $N$, the combination of our fast algorithm with modern computers 
makes long-time large-scale BD simulations with HIs within practical reach. We are currently 
incorporating these fast algorithms into the modified Metropolis integrator and applying the resulting
algorithm to study the lipid bilayer membrane of the red blood cells in the blood flow.  Results
from these ongoing work  are being analyzed now and will be reported in a timely fashion.

\section*{Acknowledgement}
Y.-N. Young was supported by NSF under grant DMS-1222550.
S. Jiang was supported by NSF under grant DMS-1418918. 
The authors would like to thank Dr. Bou-Rabee and Dr. Schroeder for helpful
discussions.

\bibliographystyle{aipauth4-1}
\bibliography{Fu2014}

%merlin.mbs aipauth4-1.bst 2010-07-25 4.21a (PWD, AO, DPC) hacked
%Control: author (9) reversed initials
\begin{thebibliography}{26}%
\makeatletter
\providecommand \@ifxundefined [1]{%
 \@ifx{#1\undefined}
}%
\providecommand \@ifnum [1]{%
 \ifnum #1\expandafter \@firstoftwo
 \else \expandafter \@secondoftwo
 \fi
}%
\providecommand \@ifx [1]{%
 \ifx #1\expandafter \@firstoftwo
 \else \expandafter \@secondoftwo
 \fi
}%
\providecommand \natexlab [1]{#1}%
\providecommand \enquote  [1]{``#1''}%
\providecommand \bibnamefont  [1]{#1}%
\providecommand \bibfnamefont [1]{#1}%
\providecommand \citenamefont [1]{#1}%
\providecommand \href@noop [0]{\@secondoftwo}%
\providecommand \href [0]{\begingroup \@sanitize@url \@href}%
\providecommand \@href[1]{\@@startlink{#1}\@@href}%
\providecommand \@@href[1]{\endgroup#1\@@endlink}%
\providecommand \@sanitize@url [0]{\catcode `\\12\catcode `\$12\catcode
  `\&12\catcode `\#12\catcode `\^12\catcode `\_12\catcode `\%12\relax}%
\providecommand \@@startlink[1]{}%
\providecommand \@@endlink[0]{}%
\providecommand \url  [0]{\begingroup\@sanitize@url \@url }%
\providecommand \@url [1]{\endgroup\@href {#1}{\urlprefix }}%
\providecommand \urlprefix  [0]{URL }%
\providecommand \Eprint [0]{\href }%
\providecommand \doibase [0]{http://dx.doi.org/}%
\providecommand \selectlanguage [0]{\@gobble}%
\providecommand \bibinfo  [0]{\@secondoftwo}%
\providecommand \bibfield  [0]{\@secondoftwo}%
\providecommand \translation [1]{[#1]}%
\providecommand \BibitemOpen [0]{}%
\providecommand \bibitemStop [0]{}%
\providecommand \bibitemNoStop [0]{.\EOS\space}%
\providecommand \EOS [0]{\spacefactor3000\relax}%
\providecommand \BibitemShut  [1]{\csname bibitem#1\endcsname}%
\let\auto@bib@innerbib\@empty
%</preamble>
\bibitem [{\citenamefont {Bou-Rabee}(2014)}]{BouRabee14}%
  \BibitemOpen
  \bibfield  {author} {\bibinfo {author} {\bibnamefont {Bou-Rabee},
  \bibfnamefont {N.}},\ }\href@noop {} {\bibfield  {journal} {\bibinfo
  {journal} {Entropy}\ }\textbf {\bibinfo {volume} {16}},\ \bibinfo {pages}
  {138} (\bibinfo {year} {2014})}\BibitemShut {NoStop}%
\bibitem [{\citenamefont {Bou-Rabee}, \citenamefont {Donev},\ and\
  \citenamefont {Vanden-Eijnden}(2014)}]{BouRabee14v2}%
  \BibitemOpen
  \bibfield  {author} {\bibinfo {author} {\bibnamefont {Bou-Rabee},
  \bibfnamefont {N.}}, \bibinfo {author} {\bibnamefont {Donev}, \bibfnamefont
  {A.}}, \ and\ \bibinfo {author} {\bibnamefont {Vanden-Eijnden}, \bibfnamefont
  {E.}},\ }\href@noop {} {\bibfield  {journal} {\bibinfo  {journal} {Multiscale
  Model. Simul.}\ }\textbf {\bibinfo {volume} {12(2)}},\ \bibinfo {pages} {781}
  (\bibinfo {year} {2014})}\BibitemShut {NoStop}%
\bibitem [{\citenamefont {Doi}\ and\ \citenamefont
  {Edwards}(1986)}]{DoiEdwards}%
  \BibitemOpen
  \bibfield  {author} {\bibinfo {author} {\bibnamefont {Doi}, \bibfnamefont
  {M.}}\ and\ \bibinfo {author} {\bibnamefont {Edwards}, \bibfnamefont
  {S.~F.}},\ }\href@noop {} {\emph {\bibinfo {title} {The Theory of Polymer
  Dynamics}}}\ (\bibinfo  {publisher} {Oxford Science Publications},\ \bibinfo
  {address} {New York},\ \bibinfo {year} {1986})\BibitemShut {NoStop}%
\bibitem [{\citenamefont {Ermak}\ and\ \citenamefont
  {McCammon}(1978)}]{Ermak78}%
  \BibitemOpen
  \bibfield  {author} {\bibinfo {author} {\bibnamefont {Ermak}, \bibfnamefont
  {D.~L.}}\ and\ \bibinfo {author} {\bibnamefont {McCammon}, \bibfnamefont
  {J.~A.}},\ }\href@noop {} {\bibfield  {journal} {\bibinfo  {journal} {J.
  Chem. Phys.}\ }\textbf {\bibinfo {volume} {69}},\ \bibinfo {pages} {1352}
  (\bibinfo {year} {1978})}\BibitemShut {NoStop}%
\bibitem [{\citenamefont {Greengard}\ and\ \citenamefont
  {Rokhlin}(1987)}]{fmm}%
  \BibitemOpen
  \bibfield  {author} {\bibinfo {author} {\bibnamefont {Greengard},
  \bibfnamefont {L.}}\ and\ \bibinfo {author} {\bibnamefont {Rokhlin},
  \bibfnamefont {V.}},\ }\href@noop {} {\bibfield  {journal} {\bibinfo
  {journal} {J. Comput. Phys.}\ }\textbf {\bibinfo {volume} {73}},\ \bibinfo
  {pages} {325} (\bibinfo {year} {1987})}\BibitemShut {NoStop}%
\bibitem [{\citenamefont {H.~Cheng}\ and\ \citenamefont
  {Rokhlin}(1999)}]{cheng}%
  \BibitemOpen
  \bibfield  {author} {\bibinfo {author} {\bibnamefont {H.~Cheng},
  \bibfnamefont {L.~G.}}\ and\ \bibinfo {author} {\bibnamefont {Rokhlin},
  \bibfnamefont {V.}},\ }\href@noop {} {\bibfield  {journal} {\bibinfo
  {journal} {J. Comput. Phys.}\ }\textbf {\bibinfo {volume} {155}},\ \bibinfo
  {pages} {468} (\bibinfo {year} {1999})}\BibitemShut {NoStop}%
\bibitem [{\citenamefont {H.~P.~Babcock}(2000)}]{Chu00}%
  \BibitemOpen
  \bibfield  {author} {\bibinfo {author} {\bibnamefont {H.~P.~Babcock},
  \bibfnamefont {D.~E.~Smith, J.~H. E. S. G. S. S.~C.}},\ }\href@noop {}
  {\bibfield  {journal} {\bibinfo  {journal} {Phys. Rev. Lett.}\ }\textbf
  {\bibinfo {volume} {85(9)}},\ \bibinfo {pages} {2018} (\bibinfo {year}
  {2000})}\BibitemShut {NoStop}%
\bibitem [{\citenamefont {Ho}\ and\ \citenamefont {Ying}(2013)}]{kenho1}%
  \BibitemOpen
  \bibfield  {author} {\bibinfo {author} {\bibnamefont {Ho}, \bibfnamefont
  {K.~L.}}\ and\ \bibinfo {author} {\bibnamefont {Ying}, \bibfnamefont {L.}},\
  }\href@noop {} {\bibfield  {journal} {\bibinfo  {journal} {arXiv:1307.2666}\
  } (\bibinfo {year} {2013})}\BibitemShut {NoStop}%
\bibitem [{\citenamefont {Hsieh}, \citenamefont {Li.},\ and\ \citenamefont
  {Larson}(2003)}]{Larson03}%
  \BibitemOpen
  \bibfield  {author} {\bibinfo {author} {\bibnamefont {Hsieh}, \bibfnamefont
  {C.-C.}}, \bibinfo {author} {\bibnamefont {Li.}, \bibfnamefont {L.}}, \ and\
  \bibinfo {author} {\bibnamefont {Larson}, \bibfnamefont {R.~G.}},\
  }\href@noop {} {\bibfield  {journal} {\bibinfo  {journal} {J. Non-Newton
  Fluid}\ }\textbf {\bibinfo {volume} {113}},\ \bibinfo {pages} {147} (\bibinfo
  {year} {2003})}\BibitemShut {NoStop}%
\bibitem [{\citenamefont {Jendrejack}, \citenamefont {Graham},\ and\
  \citenamefont {de~Pablo}(2000)}]{Graham00}%
  \BibitemOpen
  \bibfield  {author} {\bibinfo {author} {\bibnamefont {Jendrejack},
  \bibfnamefont {R.~M.}}, \bibinfo {author} {\bibnamefont {Graham},
  \bibfnamefont {M.~D.}}, \ and\ \bibinfo {author} {\bibnamefont {de~Pablo},
  \bibfnamefont {J.~J.}},\ }\href@noop {} {\bibfield  {journal} {\bibinfo
  {journal} {J. Chem. Phys.}\ }\textbf {\bibinfo {volume} {113}},\ \bibinfo
  {pages} {2894} (\bibinfo {year} {2000})}\BibitemShut {NoStop}%
\bibitem [{\citenamefont {Jendrejack}, \citenamefont {de~Pablo},\ and\
  \citenamefont {Graham}(2002)}]{Graham02}%
  \BibitemOpen
  \bibfield  {author} {\bibinfo {author} {\bibnamefont {Jendrejack},
  \bibfnamefont {R.~M.}}, \bibinfo {author} {\bibnamefont {de~Pablo},
  \bibfnamefont {J.~J.}}, \ and\ \bibinfo {author} {\bibnamefont {Graham},
  \bibfnamefont {M.~D.}},\ }\href@noop {} {\bibfield  {journal} {\bibinfo
  {journal} {J. Chem. Phys.}\ }\textbf {\bibinfo {volume} {116}},\ \bibinfo
  {pages} {7752} (\bibinfo {year} {2002})}\BibitemShut {NoStop}%
\bibitem [{\citenamefont {Jiang}, \citenamefont {Liang},\ and\ \citenamefont
  {Huang}(2013)}]{Jiang13_MC}%
  \BibitemOpen
  \bibfield  {author} {\bibinfo {author} {\bibnamefont {Jiang}, \bibfnamefont
  {S.}}, \bibinfo {author} {\bibnamefont {Liang}, \bibfnamefont {Z.}}, \ and\
  \bibinfo {author} {\bibnamefont {Huang}, \bibfnamefont {J.}},\ }\href@noop {}
  {\bibfield  {journal} {\bibinfo  {journal} {Math Comput.}\ }\textbf {\bibinfo
  {volume} {82}},\ \bibinfo {pages} {1631} (\bibinfo {year}
  {2013})}\BibitemShut {NoStop}%
\bibitem [{\citenamefont {K.~L.~Ho}(2013)}]{kenho2}%
  \BibitemOpen
  \bibfield  {author} {\bibinfo {author} {\bibnamefont {K.~L.~Ho},
  \bibfnamefont {L.~Y.}},\ }\href@noop {} {\bibfield  {journal} {\bibinfo
  {journal} {arXiv:1307.2895}\ } (\bibinfo {year} {2013})}\BibitemShut
  {NoStop}%
\bibitem [{\citenamefont {Larson}\ \emph {et~al.}(1999)\citenamefont {Larson},
  \citenamefont {Hua}, \citenamefont {Smith},\ and\ \citenamefont
  {Chu}}]{Larson99}%
  \BibitemOpen
  \bibfield  {author} {\bibinfo {author} {\bibnamefont {Larson}, \bibfnamefont
  {R.~G.}}, \bibinfo {author} {\bibnamefont {Hua}, \bibfnamefont {H.}},
  \bibinfo {author} {\bibnamefont {Smith}, \bibfnamefont {D.~E.}}, \ and\
  \bibinfo {author} {\bibnamefont {Chu}, \bibfnamefont {S.}},\ }\href@noop {}
  {\bibfield  {journal} {\bibinfo  {journal} {J. Rheol.}\ }\textbf {\bibinfo
  {volume} {46}},\ \bibinfo {pages} {267} (\bibinfo {year} {1999})}\BibitemShut
  {NoStop}%
\bibitem [{\citenamefont {Liang}\ \emph {et~al.}(2013)\citenamefont {Liang},
  \citenamefont {Gimbutas}, \citenamefont {Greengard}, \citenamefont {Huang},\
  and\ \citenamefont {Jiang}}]{Jiang13_JCP}%
  \BibitemOpen
  \bibfield  {author} {\bibinfo {author} {\bibnamefont {Liang}, \bibfnamefont
  {Z.}}, \bibinfo {author} {\bibnamefont {Gimbutas}, \bibfnamefont {Z.}},
  \bibinfo {author} {\bibnamefont {Greengard}, \bibfnamefont {L.}}, \bibinfo
  {author} {\bibnamefont {Huang}, \bibfnamefont {J.}}, \ and\ \bibinfo {author}
  {\bibnamefont {Jiang}, \bibfnamefont {S.}},\ }\href@noop {} {\bibfield
  {journal} {\bibinfo  {journal} {J. Comput. Phys.}\ }\textbf {\bibinfo
  {volume} {234}},\ \bibinfo {pages} {133} (\bibinfo {year}
  {2013})}\BibitemShut {NoStop}%
\bibitem [{\citenamefont {Marko}\ and\ \citenamefont {Siggia}(1995)}]{Marko95}%
  \BibitemOpen
  \bibfield  {author} {\bibinfo {author} {\bibnamefont {Marko}, \bibfnamefont
  {J.~F.}}\ and\ \bibinfo {author} {\bibnamefont {Siggia}, \bibfnamefont
  {E.~D.}},\ }\href@noop {} {\bibfield  {journal} {\bibinfo  {journal}
  {Macromolecules}\ }\textbf {\bibinfo {volume} {28}},\ \bibinfo {pages} {8759}
  (\bibinfo {year} {1995})}\BibitemShut {NoStop}%
\bibitem [{\citenamefont {Perkins}, \citenamefont {Smith},\ and\ \citenamefont
  {Chu}(1997)}]{Chu97}%
  \BibitemOpen
  \bibfield  {author} {\bibinfo {author} {\bibnamefont {Perkins}, \bibfnamefont
  {T.~T.}}, \bibinfo {author} {\bibnamefont {Smith}, \bibfnamefont {D.~E.}}, \
  and\ \bibinfo {author} {\bibnamefont {Chu}, \bibfnamefont {S.}},\ }\href@noop
  {} {\bibfield  {journal} {\bibinfo  {journal} {Science}\ }\textbf {\bibinfo
  {volume} {276}},\ \bibinfo {pages} {2016} (\bibinfo {year}
  {1997})}\BibitemShut {NoStop}%
\bibitem [{\citenamefont {Prakash}(2002)}]{Prakash02}%
  \BibitemOpen
  \bibfield  {author} {\bibinfo {author} {\bibnamefont {Prakash}, \bibfnamefont
  {J.~R.}},\ }\href@noop {} {\bibfield  {journal} {\bibinfo  {journal} {J.
  Rheol.}\ }\textbf {\bibinfo {volume} {46}},\ \bibinfo {pages} {1353}
  (\bibinfo {year} {2002})}\BibitemShut {NoStop}%
\bibitem [{\citenamefont {Rotne}\ and\ \citenamefont {Prager}(1969)}]{RP69}%
  \BibitemOpen
  \bibfield  {author} {\bibinfo {author} {\bibnamefont {Rotne}, \bibfnamefont
  {J.}}\ and\ \bibinfo {author} {\bibnamefont {Prager}, \bibfnamefont {S.}},\
  }\href@noop {} {\bibfield  {journal} {\bibinfo  {journal} {J. Chem. Phys.}\
  }\textbf {\bibinfo {volume} {50}},\ \bibinfo {pages} {4831} (\bibinfo {year}
  {1969})}\BibitemShut {NoStop}%
\bibitem [{\citenamefont {S.~Ambikasaran}\ and\ \citenamefont
  {O'Neil}(2014)}]{siba}%
  \BibitemOpen
  \bibfield  {author} {\bibinfo {author} {\bibnamefont {S.~Ambikasaran},
  \bibfnamefont {D.~Foreman-Mackey, L.~G. D. W.~H.}}\ and\ \bibinfo {author}
  {\bibnamefont {O'Neil}, \bibfnamefont {M.}},\ }\href@noop {} {\bibfield
  {journal} {\bibinfo  {journal} {Preprint}\ } (\bibinfo {year}
  {2014})}\BibitemShut {NoStop}%
\bibitem [{\citenamefont {Schroeder}(2014)}]{SchroederPrivateComm}%
  \BibitemOpen
  \bibfield  {author} {\bibinfo {author} {\bibnamefont {Schroeder},
  \bibfnamefont {C.~M.}},\ }\href@noop {} {\bibfield  {journal} {\bibinfo
  {journal} {(private communication)}\ } (\bibinfo {year} {2014})}\BibitemShut
  {NoStop}%
\bibitem [{\citenamefont {Schroeder}\ \emph {et~al.}(2003)\citenamefont
  {Schroeder}, \citenamefont {Babcock}, \citenamefont {Shaqfeh},\ and\
  \citenamefont {Chu}}]{Schroeder03}%
  \BibitemOpen
  \bibfield  {author} {\bibinfo {author} {\bibnamefont {Schroeder},
  \bibfnamefont {C.~M.}}, \bibinfo {author} {\bibnamefont {Babcock},
  \bibfnamefont {H.~P.}}, \bibinfo {author} {\bibnamefont {Shaqfeh},
  \bibfnamefont {E.~S.~G.}}, \ and\ \bibinfo {author} {\bibnamefont {Chu},
  \bibfnamefont {S.}},\ }\href@noop {} {\bibfield  {journal} {\bibinfo
  {journal} {Science}\ }\textbf {\bibinfo {volume} {301}},\ \bibinfo {pages}
  {1515} (\bibinfo {year} {2003})}\BibitemShut {NoStop}%
\bibitem [{\citenamefont {Schroeder}, \citenamefont {Shaqfeh},\ and\
  \citenamefont {Chu}(2004)}]{Schroeder04}%
  \BibitemOpen
  \bibfield  {author} {\bibinfo {author} {\bibnamefont {Schroeder},
  \bibfnamefont {C.~M.}}, \bibinfo {author} {\bibnamefont {Shaqfeh},
  \bibfnamefont {E.~S.~G.}}, \ and\ \bibinfo {author} {\bibnamefont {Chu},
  \bibfnamefont {S.}},\ }\href@noop {} {\bibfield  {journal} {\bibinfo
  {journal} {Macromolecules}\ }\textbf {\bibinfo {volume} {37}},\ \bibinfo
  {pages} {9242} (\bibinfo {year} {2004})}\BibitemShut {NoStop}%
\bibitem [{\citenamefont {Smith}, \citenamefont {Babcock},\ and\ \citenamefont
  {Chu}(1999)}]{Chu99}%
  \BibitemOpen
  \bibfield  {author} {\bibinfo {author} {\bibnamefont {Smith}, \bibfnamefont
  {D.~E.}}, \bibinfo {author} {\bibnamefont {Babcock}, \bibfnamefont {H.~P.}},
  \ and\ \bibinfo {author} {\bibnamefont {Chu}, \bibfnamefont {S.}},\
  }\href@noop {} {\bibfield  {journal} {\bibinfo  {journal} {Science}\ }\textbf
  {\bibinfo {volume} {283}},\ \bibinfo {pages} {1724} (\bibinfo {year}
  {1999})}\BibitemShut {NoStop}%
\bibitem [{\citenamefont {Smith}\ and\ \citenamefont {Chu}(1998)}]{Chu98}%
  \BibitemOpen
  \bibfield  {author} {\bibinfo {author} {\bibnamefont {Smith}, \bibfnamefont
  {D.~E.}}\ and\ \bibinfo {author} {\bibnamefont {Chu}, \bibfnamefont {S.}},\
  }\href@noop {} {\bibfield  {journal} {\bibinfo  {journal} {Science}\ }\textbf
  {\bibinfo {volume} {281}},\ \bibinfo {pages} {1335} (\bibinfo {year}
  {1998})}\BibitemShut {NoStop}%
\bibitem [{\citenamefont {Somasi}\ \emph {et~al.}(2002)\citenamefont {Somasi},
  \citenamefont {Khomami}, \citenamefont {Woo}, \citenamefont {Hur},\ and\
  \citenamefont {Shaqfeh}}]{Shaqfeh02}%
  \BibitemOpen
  \bibfield  {author} {\bibinfo {author} {\bibnamefont {Somasi}, \bibfnamefont
  {M.}}, \bibinfo {author} {\bibnamefont {Khomami}, \bibfnamefont {B.}},
  \bibinfo {author} {\bibnamefont {Woo}, \bibfnamefont {N.~J.}}, \bibinfo
  {author} {\bibnamefont {Hur}, \bibfnamefont {J.~S.}}, \ and\ \bibinfo
  {author} {\bibnamefont {Shaqfeh}, \bibfnamefont {E.~S.~G.}},\ }\href@noop {}
  {\bibfield  {journal} {\bibinfo  {journal} {J. Non-Newtonian Fluid Mech.}\
  }\textbf {\bibinfo {volume} {108}},\ \bibinfo {pages} {227} (\bibinfo {year}
  {2002})}\BibitemShut {NoStop}%
\end{thebibliography}%

\end{document}